\documentclass{aa}  
\usepackage{graphicx}
\usepackage{txfonts}

\usepackage{amsmath}
\usepackage{amsfonts}
\usepackage{subcaption}
\usepackage{placeins}
\usepackage{hyperref}
\usepackage{cleveref}
\usepackage{bm}
\usepackage{makecell}
\usepackage{siunitx}
\usepackage{graphicx}
\usepackage{caption}
\usepackage[normalem]{ulem}
\usepackage{tcolorbox}
\usepackage[dvipsnames]{xcolor}

\usepackage{tikz}
\usetikzlibrary{shapes.geometric, arrows, positioning}
\usetikzlibrary{backgrounds}

\usepackage{natbib}
\bibpunct{(}{)}{;}{a}{}{,}

\newcommand{\KL}{\text{KL}}
\newcommand{\E}{\mathbb{E}}

\newcommand{\bigcdot}{\mathbin{\smash{\vcenter{\hbox{\scalebox{1.3}{$\cdot$}}}}}}

\usepackage{tikz,xcolor}

\newif \ifhighlightchanges
\highlightchangesfalse

\begin{document}

\title{Bayesian classification of astronomical spectra \\\ with class uncertainties}

\author{Simon Barton \inst{1}
    \and
    Martin Sahlén \inst{1}
    \and
    Andreas Korn \inst{2}
    \and
    Christian Glaser \inst{3,4}
}

\institute{Theoretical Astrophysics, Division of Astronomy and Space Physics, Department of Physics and Astronomy, Uppsala University, Box 524, 751 20 Uppsala, Sweden. Corresponding author \email{simon.barton@physics.uu.se}
 \and 
Observational Astrophysics, Division of Astronomy and Space Physics, Department of Physics and Astronomy, Uppsala University, Box 524, 751 20 Uppsala, Sweden
 \and 
Division of High Energy Physics, Department of Physics and Astronomy, Uppsala University, Box 524, 751 20 Uppsala, Sweden
\and
Dept. of Physics, TU Dortmund University, D-44221 Dortmund, Germany
}

\abstract
{We developed a probabilistic machine learning method with the aim of performing the $\mathcal{O}(10)$-way classification of low- and high-resolution spectra of stellar and extragalactic targets for the upcoming 4MOST survey.
In fulfilment of the survey requirements, this method should be able to express uncertainty in the input data as well as uncertainty introduced in its prediction.
}
{Four different methods are explored: (1) convolutional neural networks (CNNs), (2) the Dirichlet distribution, (3) Monte Carlo dropout (MCD), (4) Bayesian neural Networks (BNNs) + variational inference (VI).
    Training and validation was performed using labelled spectra from  the SDSS database and a custom 4MOST mock dataset.
    All the methods were compared in terms of the same metrics: accuracy, area under the curve (AUC), expected calibration error (ECE), Shannon entropy, negative log-likelihood (NLL), Brier score, training time, and inference time.
}
{A CNN with simple architecture and $\sim2\times 10^5$ parameters was trained to achieve classification accuracies of $91.5\%$ on SDSS data and $92.8\%$ on 4MOST mock data. 
The direct Dirichlet prediction and VI models tested provide uncertainties on class membership probabilities, but they confuse classes more often.
The MCD on a CNN is found to be the most suitable; it boosts the point-estimate accuracies to $92.6\%$ and $93.9\%$, while still providing fast training and sufficiently fast inference. Compared to a standard CNN, the method additionally provides well-calibrated uncertainties at marginal extra cost. 
}
{}


\maketitle
\nolinenumbers 

\section{Introduction}

The next-generation 4-metre Multi-Object Spectroscopic Telescope (4MOST), due to start full science operations in 2026, will operate with low- and high-resolution spectrographs in parallel to observe up to 2436 celestial targets simultaneously \citep{de20194most}.\footnote{\url{https://www.4most.eu}}
Over its first five years, the instrument will acquire up to 40 million spectra, necessitating scalable data processing methods beyond the capabilities of template-based pipelines.
Processing the large amounts of data that come with modern astronomical spectroscopic surveys includes the classification of targets, both for confirming catalogue labels and for part of a follow-up after a first photometric detection.
Next to conventionally used line-fitting algorithms, which are slow and require physical modelling, machine learning (ML) methods have been successfully used in this processing \citep{zhong2023galaxy}.

However, such methods can lead to overconfident predictions.
Hence, the fully automated 4MOST classification pipeline calls for probabilistic models that estimate uncertainty on class membership.
Understanding and quantifying the uncertainty in classification tasks has become increasingly important, particularly in high-stakes applications where model confidence must be both interpretable and well calibrated.
The foundational notion of calibration was introduced by \citet{guo2017calibration} and was refined by the agreement between predicted probabilities and observed frequencies \citep{nixon2019measuring},
by inspired recalibration strategies \citep{kranzlein2021making}, 
and by improved evaluation techniques for multi-class settings \citep{silva2023classifier}.

In parallel, the literature has expanded on how to represent and decompose predictive uncertainty. The methods range from geometric and entropy-based measures \citep{chlaily2023measures} to Bayesian approaches that consider the distribution over performance metrics derived from confusion matrices rather than predictive distributions directly \citep{caelen2017bayesian}.

Bayesian neural networks (BNNs) have emerged as promising approaches for modelling uncertainty in deep learning by placing probability distributions over model parameters.
Unlike conventional neural networks that yield point estimates, BNNs capture epistemic uncertainty by marginalizing over these distributions, thus resulting in predictive probabilities that reflect both model confidence and data ambiguity. This makes them particularly well suited  for applications where calibrated uncertainty estimates are critical, such as classification under limited or noisy data \citep{gawlikowski2023survey, goan2020bayesian}.

The decomposition of uncertainty into aleatoric and epistemic components was implemented in practical deep learning settings using Bayesian deep learning models \citep{kendall2017uncertainties, mancarella2022seeking}.
It has been shown how proper scoring rules can be used to quantify and distinguish between these two types of uncertainty in a principled manner \citep{hofman2024quantifying}.
BNNs have been applied in many fields, such as gravitational waves \citep{lin2021detection} and the cosmic microwave background,
\citep{hortua2020parameter} and used to estimate redshifts of galaxies from photometric DESI data \citep{zhou2025estimating}.
Specifically for classification, they have been compared to other Bayesian deep learning classifiers for supernova signals \citep{moller2020supernnova} and to assess modified theories of gravity \citep{thummel2024classifying}.
The comparisons have been made between different methods for uncertainty estimation and their robustness against out-of-distribution data \citep{vranken2021uncertainty, mancarella2022seeking}.

All methods were implemented using PyTorch \citep{paszke2019pytorch}.
The entire machine learning code is available online.\footnote{\url{https://codeberg.org/simonbarton/bnn}}
Throughout this work the notation $|X|$ is used for the number of elements in $X$.

\section{Spectroscopic datasets}
To test our classification algorithms we used two custom datasets of low-resolution optical spectra.
Both datasets were created for a spectrum classification task in \citet{zhong2023galaxy}.
To remove distance-dependent variations and bring the data into a form suitable for deep learning, the unnormalized fluxes were rescaled such that the sum of all flux values was equal to 1000 for each spectrum. We note that the first BatchNorm layer normalizes the fluxes again.
In total, we used 70\% spectra for training and 15\% for validation and testing, respectively.
The validation set was used to select the hyperparameters and perform early stopping, while the test set was used to compare the results across different models.
All splits were artificially class-balanced to ensure stable training and fair comparison across classes. 
This allowed us to better assess the model’s ability to distinguish minority classes and also to ensure comparability to other models.
However, when applied within 4MOST, the true class distribution will be skewed and using a test set that reflects the actual astronomical object abundances may provide a more realistic estimate of deployment performance.
In particular, the models may overestimate performance on rare classes and underestimate overall uncertainty under natural class priors.
An estimate of the final accuracy for the SDSS dataset is given in Sect. \ref{sec:results}, assuming the target class balance matches the SDSS catalogue abundance.

\subsection{SDSS}
The spectra of our first dataset were obtained from eBOSS and are available as part of the SDSS Data Release 16 \citep{jonsson2020apogee}.

\begin{figure}[h]
    \centering
    \includegraphics[width=\hsize]{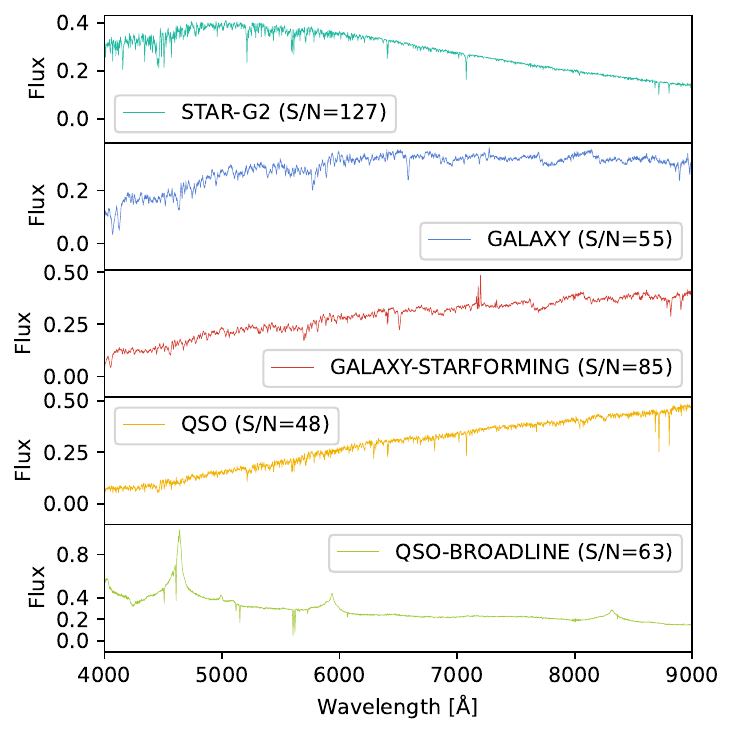}
    \caption{
    Selected example spectra of the SDSS dataset with the highest S/N of the respective class.
    }
    \label{fig:data_spectra}
\end{figure}

Unnormalized fluxes are given for 3,600 wavelengths between $4000Å$ and $9000Å$.
The labels correspond to the 13 SDSS (sub)classes with more than 20 000 classified objects: seven galactic classes of different main sequence stars (distinguished by their surface temperature according to the Harvard spectral classification system, where each letter is further subdivided numerically, e.g. G2 star) and six extragalactic classes (four galaxies of different star-forming activity and nucleus activity, plus quasars with and without broad lines).
Although the selected classes only constitute a small subset of the total 181 SDSS subclasses, they are similar to the planned 4MOST classification labels in their coverage of a wide range of both galactic and extragalactic objects.
In total, we used 260,000 spectra in splits, as described above.
Some example spectra are displayed in Fig. \ref{fig:data_spectra}.

\begin{figure}[h]
    \centering
    \includegraphics[width=\hsize]{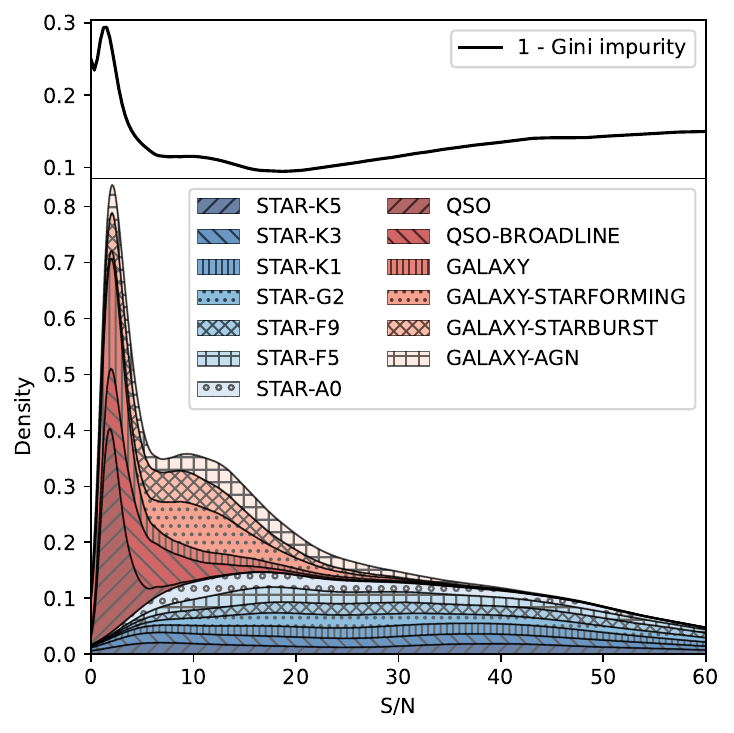}
    \caption{
        S/N distribution of each class for training and validation sets (bottom).
        Galactic (blue) and extragalactic (red) targets dominate the low- and high-noise regimes, respectively.
        Expected random-guessing accuracy based on the class ratio (top). For low S/N, the labels become easier to predict.
    }
    \label{fig:data_entropy}
\end{figure}
In addition, class balance   varies with the signal-to-noise ratio (S/N; see Fig. \ref{fig:data_entropy}).
Even though the validation set is a good representation of the total training data, this correlation can affect classification behaviour across the S/N range.
To quantify the effect, we calculated the Gini purity $G = \sum_{i=1}^{K} p_i^2$ with the number of classes $K=13$.
In the context of classification, $G$ provides an estimate for the expected accuracy of a classifier that guesses labels based solely on the class distribution, without considering the input spectrum.
For highly imbalanced datasets, where one class dominates, $G \approx 1$ leads to a high baseline accuracy.
Conversely, balanced datasets where class probabilities are uniform lead to $G = 1/K$, corresponding to the expected accuracy of random guessing.

\subsection{4MOST mock}

As a second dataset we used around 200,000 spectra in the same splits as above, simulated using the 4MOST Exposure Time Calculator based on templates for selected 4MOST surveys.
The optical wavelength range from $4000Å$ and $9000Å$ and the wavelength resolution of $5001$ are comparable to the products of the 4MOST low-resolution spectrographs.

The five galactic classes from the 4MOST surveys S1--S4 and S9 are metal-poor stars and other dynamics tracers (Dyn), Cepheids in the Magellanic Clouds (GalHR), white dwarfs (ESN), and stars of the galactic disc (GalDiskLR) and of the Magellanic Clouds (MCsn).
The five extragalactic classes  represent  the targets of the extragalactic 4MOST surveys S5--S8 and S10.
In particular, we have S6 active galactic nuclei (AGN; COSMO\textunderscore AGN), S5 bright cluster galaxies (ClusB), S7 galaxies (WAVES), S8 red galaxies (RedGAL), and S10 supernova host galaxies (tides\textunderscore host).
Example spectra for some classes and the S/N distribution of this dataset are shown in Figs. \ref{fig:mock_data_spectra} and \ref{fig:mock_data_entropy}, respectively.

\section{Uncertainty in classification} \label{sec:uncertainties}

\begin{figure*}[t!]
    \centering
    \begin{subfigure}{0.33\hsize}
        \includegraphics[width=\hsize]{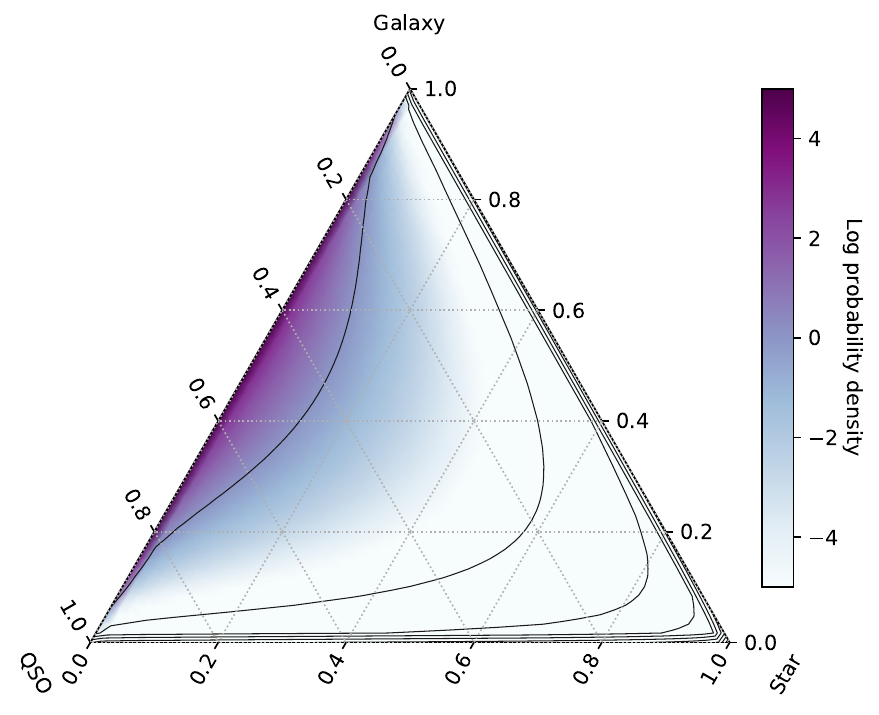}
    \end{subfigure}%
    \begin{subfigure}{0.33\hsize}
        \includegraphics[width=\hsize]{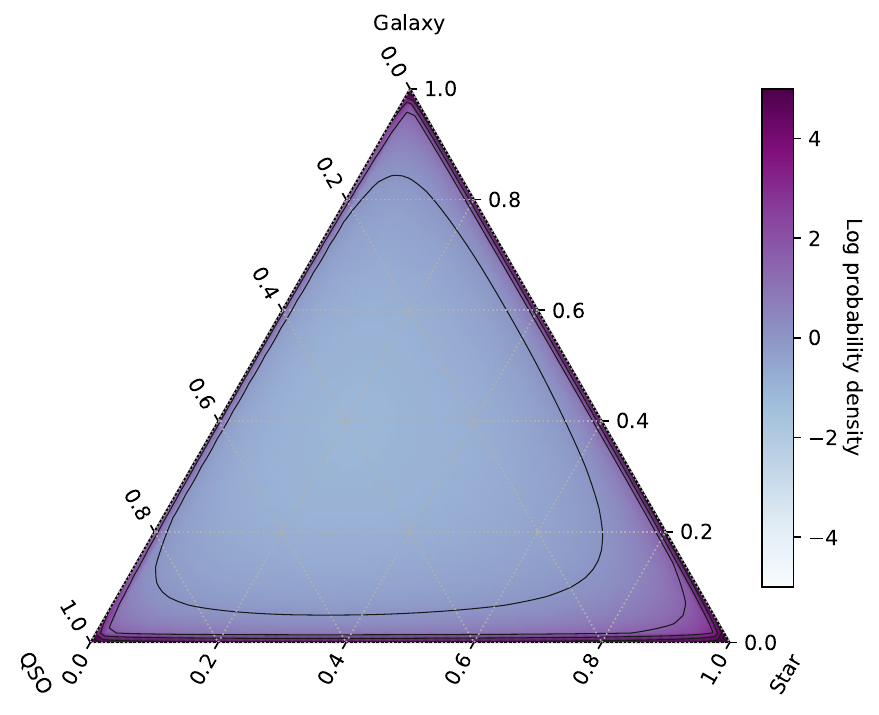}
    \end{subfigure}%
    \begin{subfigure}{0.33\hsize}
        \includegraphics[width=\hsize]{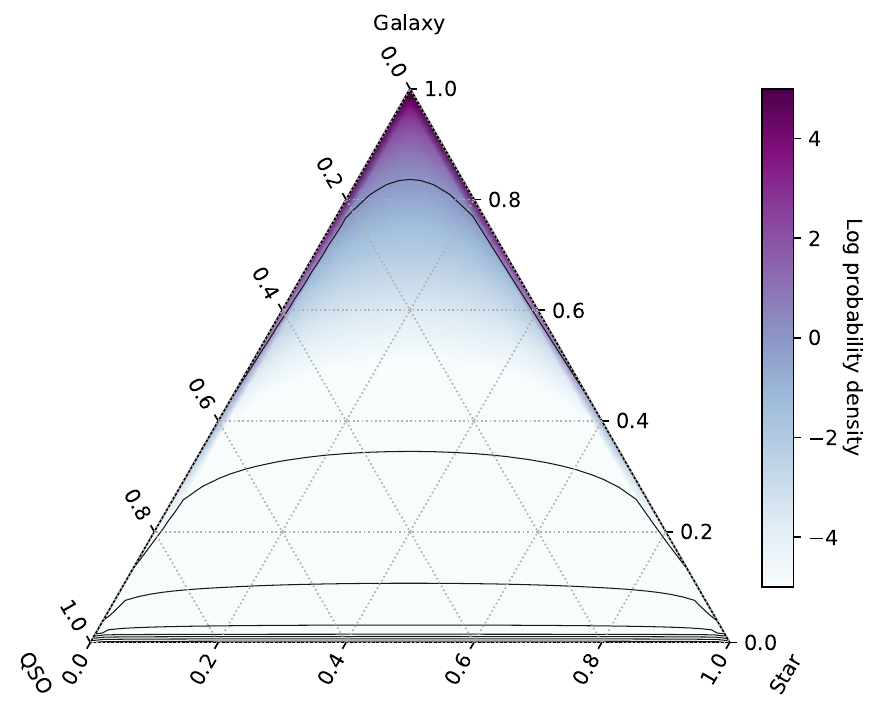}
    \end{subfigure}
    
    \caption{    
        Demonstration of the entropies of a predictive probability distribution in the three-simplex. 
        Left: Large Shannon entropy $H$ and large differential entropy $h$. The model assigns a high aleatoric ambiguity between galaxy and QSO to the input spectrum.  
        Centre: Large $H$ and small $h$. The model assigns low aleatoric uncertainty to the input spectrum, but is epistemically uncertain about its own prediction. 
        Right: Small $H$ and $h$. The model is certain about the spectrum and its prediction.
    }
    \label{fig:contours}
\end{figure*}
Given that perfect classification of these spectra is generally unachievable, for example due to noise or imprecise labels, providing uncertainties of the predictions is essential.
To formalize the problem, we denote the data as $D=\{x_i,y_i\}_{i=1}^{N}$ where each $x_i$ is a spectrum and each $y_i \in \{1,\dots,K\}$ the true label.
A non-probabilistic classifier $m$ is then a function $m:x_i\mapsto p^*_{i\bigcdot}$, where $p^*_{ik}$ is the confidence that the i-th sample belongs to the k-th class and $\bigcdot$ acts as a placeholder.
The predicted class $y^*_i = \text{argmax}_k(\{p^*_{ik}\})$ is simply the one with the highest confidence assigned.
The canonical metric to evaluate such a model is the accuracy,
\begin{align}
    \text{accuracy} = \frac{1}{N}\sum_i^N \delta_{y^*_i=y_i} \quad ,
\end{align}
with $\delta_{a=b} = 1$ if $a=b$ and $0$ otherwise.
Complementary insight is given by the receiver operating characteristic (ROC) as the relation the true positive rate (efficiency) against the false positive rate (contamination), or in summary its integral (AUC).
An ideal classifier achieves an AUC of $1$, whereas a model making random predictions yields an AUC of $0.5$.
While widely used for evaluating binary classification models, we use it in a one-versus-rest fashion for this multi-class problem.

\subsection{Calibration}
To measure the quality of the vector $p^*_{\bigcdot \bigcdot}$ directly, we compute the calibration.
A model $m$ is said to be calibrated exactly if \citep{guo2017calibration}
\begin{equation}
    \mathbb{P}( y^* = y \mid p^*_{iy^*} = p) = p \quad \forall p\in[0,1] \quad \forall i\in\{1,\dots,N\}  \quad .
\end{equation}
For example, the model's class membership predictions with a confidence of $p^*=30\%$ should coincide with the true label in $30\%$ of the cases.
The discrepancy between the predicted confidence and the empirical accuracy is summarized in the expected calibration error (ECE), which in the limit of infinitely many bins\footnote{The ECE can significantly increase with the chosen number of bins, which is why we report only the asymptotic limit.} reads
\begin{align}
    \text{ECE} = \frac{1}{K}\sum_k^K \text{ECE}_k ~, \quad \text{ECE}_k = \frac{1}{N}\sum_{i}^{N} \Big| \delta_{y^*_i=y_i} - p^*_{ik} \Big| \quad .
\end{align}
The ECE (or related error definitions) is essential for any classifier. It is a low calibration error and not the construction via a softmax function, which justifies  identifying the output confidence values $p^*$ with probabilities.

\subsection{Entropy}

One possibility to quantify the uncertainty of a single point estimate is the Shannon entropy
\begin{align}
    H(p^*_{i\,\bigcdot}) := -\sum_k^K p^*_{ik} \log p^*_{ik}  \quad \text{with} \quad 0 \log 0 := 0 \quad ,
\end{align}
which measures the distribution of confidence across classes. 
Model $m$ remains a point-estimate predictor, in that it is deterministic and it provides a single categorical distribution $p^*$. 
Probabilistic models instead predict a second-order distribution $p^*_{i}(y)$ over these $p^*_{i\bigcdot}$ \citep{gawlikowski2023survey}.
The set of all point estimates can be identified with the $(K-1)$-standard simplex $S_{k-1}$, whose $K$ corners correspond to the classes, and each convex combination thereof represents one categorical distribution with normalized probability mass (see Fig. \ref{fig:contours}). 
This allows the calculation of the differential entropy of the joint simplex distribution:
\begin{align}
    h(p^*_{i}) = - \int_{S_{k-1}} p^*_{i}(y) \log(p^*_{i}(y)) \, dy \quad .
\end{align}
Unfortunately, in high dimensions it is practically unfeasible to calculate this entropy from samples without assumptions about the form of the underlying distribution.
We tried various methods, such as kernel density estimation, without satisfying results, and therefore do not use the joint entropy as an uncertainty estimate in our discussion.

The uncertainty of a prediction $y^*_i$ comprises two distinct components.
Aleatoric uncertainty (AU) arises from inherent random noise in the data-generating process and is irreducible, regardless of the amount of data available.
In contrast, epistemic uncertainty (EU) reflects the model’s ignorance due to limited data or limited model capacity, and can be reduced by incorporating additional information, either in the form of model architecture or increasing the training data.
The total predicted uncertainty of a single output $p^*(y)$ can be decomposed into these two components \citep{depeweg2018decomposition} as $h(p(y \mid x)) = \text{AU} + \text{EU}$ with 
\begin{align}
    \text{AU} = \mathbb{E}_{\omega}[h(p^*(y \mid x, \omega))]~, \qquad \text{EU} = H(\mathbb{E}_{\omega}[p^*(y \mid x, \omega)])~,
\end{align}
where $p^*$ denotes the posterior, $h$ the entropy of the predictive distribution, and $\omega$ the latent model parameters.
Visually, this separation corresponds to the fact that an epistemic uncertainty translates to the flatness of the predicted simplex distribution, whereas aleatoric uncertainty is represented by a sharp but centred distribution \citep{malinin2018predictive, gawlikowski2023survey}.
From this it follows that the Shannon entropy specifies aleatoric uncertainty, and any point estimator is by design unaware of any epistemic uncertainty because it assumes precise knowledge of the predictive distribution \citep{gawlikowski2023survey} (see Fig. \ref{fig:contours} for the different behaviours in a three-class case).

\begin{figure*}[t!]
    \centering
    \begin{subfigure}{0.95\hsize}
        \includegraphics[width=\hsize]{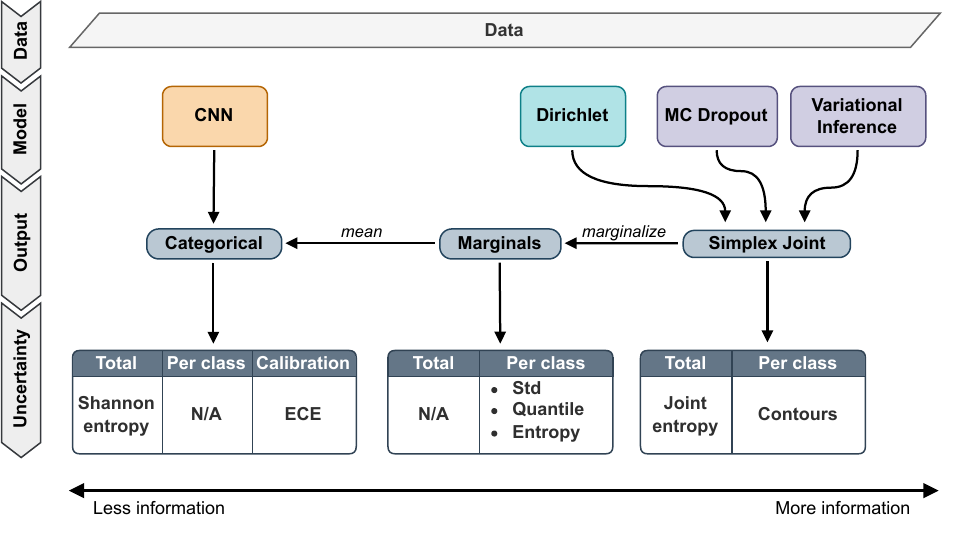}
    \end{subfigure}
    \caption{Overview scheme for this study. The arrows indicate flow of data.}
    \label{fig:overview}
\end{figure*}

\subsection{Proper scoring rules}
Another rigorous formulation of calibrated uncertainties is provided by proper scoring rules (see \citet{lakshminarayanan2017simple} and \citet{gal2016uncertainty} for a definition and discussion).
They are already used during the training in the form of the cross-entropy, and guarantee there that the point estimates are calibrated. The same principle can be applied to the continuous second-order distribution of probabilistic models.
Two proper scores are the negative log-likelihood of the label in the joint distribution, and the mean Brier score:
\begin{align}
    \text{NLL}_i = -\log p^*_{iy_i} ~,\quad  \text{Brier}_i = \frac{1}{K} \sum_{k}^{K} \left( \delta_{y_i=k} - p^*_{ik} \right)^2 \quad .
\end{align}
Theoretically founded on the proper scoring rules, these two quantities are good measures for coherent and well-calibrated uncertainty.

\section{Methodology}

In this section we present one point-estimate neural network and three probabilistic methods for the classification of astronomical spectra: (1) by fixing a closed-form (Dirichlet) distribution over the label space, (2) by using Monte Carlo dropout (MCD) on a neural network to produce stochastic outputs, and (3) by learning the distribution over model weights that propagate into a predictive distribution using Bayesian variational inference.   
As discussed in Sect. \ref{sec:uncertainties}, we calculated different quantities to measure the prediction uncertainty.
An overview of these methods and the flow of data to obtain uncertainties from their different output types is shown in Fig. \ref{fig:overview}.

The reference convolutional neural network (CNN) and MCD are trained with cross-entropy loss, the Dirichlet model uses negative log-likelihood loss, and the VI models maximize the evidence lower bound (ELBO).
For all models, we used the Adam optimizer with weight-decay regularization \citep{loshchilov2017decoupled} and custom annealing learning rate scheduling plus early stopping based on validation loss.
If prior knowledge about the true class distribution should be reflected in the predictions, the predictive posterior can be adjusted during post-processing using a second level of Bayesian inference to incorporate the desired class prior.

\subsection{Baseline CNN} \label{sec:cnn}

\begin{figure}[h!]
    \centering
    \begin{subfigure}{\hsize}
        \includegraphics[width=\hsize]{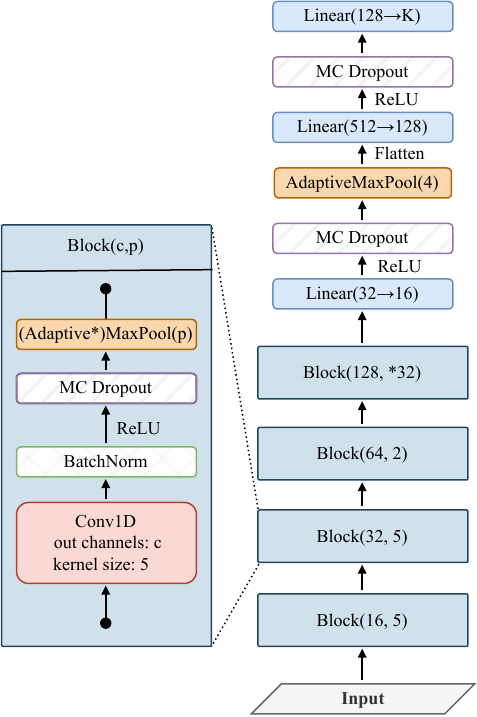}
    \end{subfigure}%
    \caption{Baseline CNN architecture, underlying all models.
    The hatched layers are enabled depending on the concrete model.
    The last block uses adaptive pooling to make the model compatible with different input spectra sizes.}
    \label{fig:cnn}
\end{figure}

As a reference model, we present a CNN for the point-estimate classification of spectra. One of the goals was to keep the model small in terms of parameters, also in view of the possibility of building more computationally demanding models based on it.
This CNN uses four blocks, where in each block a convolutional layer is followed by a batch normalization layer and a max-pooling layer to reduce the input signal (see Fig. \ref{fig:cnn}). No dropout is used for the CNN.
This model will serve as a baseline and the building block for other methods.
It has $\sim122,000$ trainable parameters, most of which  are in the dense layer head.

\subsection{Dirichlet distribution}
Because the Dirichlet distribution is the conjugate prior of the categorical distribution, it is a natural distribution over point estimates on the $(K-1)$ simplex.
Our Dirichlet model uses the baseline CNN, followed by a softplus activation to predict the concentration parameters of a $K$-dimensional Dirichlet distribution.
The loss function during training is then the negative log-likelihood of the training labels in the distribution.
For numerical reasons we offset these labels from the simplex corners,
\begin{align}
    y_i \mapsto p_{ik} = \frac{\delta_{k=y_i}+s}{1+sK},
\end{align}
and we chose the softening factor $s=10^{-4}$.

The point-estimate class membership probabilities $y^*$ can be retrieved using a Monte Carlo integration, which reduces to a simple sampling from the PDF. It is handy to embed the standard simplex into the $K$-dimensional Euclidean space such that it is spanned between the points $\{(Id_K)_i\}$, thus making the vector components correspond to the label probabilities and ensuring the independence of classes.
All other quantities can be computed similarly from these samples.

This is the only model where we were able to compute the joint entropy analytically, to represent epistemic uncertainty. However, we found this entropy not in good correlation with prediction success.

\subsection{Monte Carlo dropout}

As shown by \citet{gal2016dropout}, the Bayesian predictive posterior $p(y\mid x,\omega)$ of a neural network can be numerically approximated by using a stochastic parameter dropout.
This technique, known as Monte Carlo dropout (MCD), provides an easy-to-implement and computationally inexpensive way to combine Bayesian inference with neural networks. 

We start with the same CNN architecture as the baseline model, insert dropout layers after each batch norm operation as shown in Fig. \ref{fig:cnn}, and perform the training as usual.
During inference we then keep sampling a new dropout mask for each layer and each $x$ to obtain a distribution of predictions, represented by a collection of samples.
From there the further processing is done as for the Dirichlet method.

\begin{figure}[t!]
    \centering
    \includegraphics[width=\hsize]{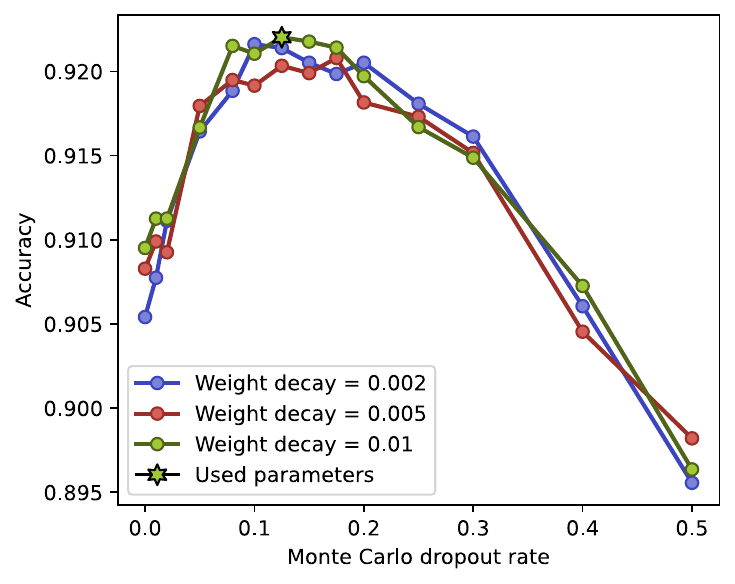}
    \caption{
        Influence of the hyperparameters dropout rate and Adam weight decay on MCD accuracy (SDSS dataset).
    }
    \label{fig:mcd_grid}
\end{figure}

The success of the method can be sensitive to the choice of hyperparameters, notably the dropout rate $r_d$ and the parameter decay rate $d_w$ of the optimizer \citep{gal2016dropout}.
In a 2D grid search (see Fig. \ref{fig:mcd_grid}) we observed a slight influence of the decay rate and a clear maximum of the accuracy for dropout rates between 8\% and 18\%. For further analysis, we selected the combination $r_d = 12.5\%, ~d_w=0.01$, which maximizes accuracy.

\subsection{Variational inference}

Extending the notation above, we now consider a probabilistic model
\begin{align}
    m_\omega : x_i \mapsto p^*_i(y) \quad ,
\end{align}
where $\omega$ denotes the internal trainable weights of $m$.
We are interested in finding the weights that explain the data best, or formally, the posterior $p(\omega\mid D)$.
Directly applying Bayes' rule,
\begin{align}
    p(\omega \mid D) = \frac{p(D \mid \omega) \; p(\omega)}{p(D)} \quad ,
\end{align}
one finds the marginal likelihood $p(D)$ to be intractable when $\omega$ is high-dimensional.
Instead, we resort to finding a surrogate $q_\theta(\omega)$, parametrized by $\theta$ to approximate the posterior.
In other words, we are looking for the value of $\theta$ that minimizes the Kullback-Leibler (KL) divergence to the true posterior,
\begin{align}
    q^*(\omega) = \text{argmin}_{\theta} \Big\{ \KL(q_\theta(\omega)\;||\;p(\omega\mid D)) \Big\} \quad .
\end{align}
From here one can rewrite the unknown posterior in terms of the known joint,
\begin{align}\label{eq:vi}
    \KL \left(q_\theta(\omega)\;\mid\mid \;p(\omega\mid D)\right) = 
    \E_{\omega \sim q_\theta}\left[ \log \frac{q_\theta(\omega) \, p(D)}{p(\omega,D)} \right] \quad .
\end{align}
In this form, the intractability has been moved into $p(D)$, but because this term is not dependent on $\theta$, the remaining expressions can be computed to form a loss function that is used in a gradient descent algorithm on $\theta$.
Once the model is trained, one can easily compute the predictive posterior from the parameter surrogate by applying Bayes' rule again,
\begin{align}
    p^*(y \mid x,\omega) = \frac{p(y \mid \omega) \, p(x,y \mid \omega)}{p(x,\omega)} \quad .
\end{align}
This method for approximating the posterior by a surrogate is called variational inference (VI).
In its stochastic version (SVI), it scales well and is therefore particularly useful in large-scale problems \citep{kingma2017variational, goan2020bayesian, gawlikowski2023survey}.

In practice, to design a BNN one chooses a functional model $m_\omega$, a stochastic model that contains both priors $p(\omega)$ and $p(y \mid x,\omega)$ and, in the case of VI, the form of the parameter posterior $p(\omega \mid D)$. This choice is the equivalent of a loss function in point-estimate ML \citep{jospin2022hands}.

For the functional model we reused our baseline CNN. 
As a fixed form of $q_\theta$ we chose the multivariate normal distribution with low-rank covariance.
As done in \citet{ong2018gaussian} we factorize the covariance matrix
\begin{align}
    \Sigma = BB^T + C^2
\end{align}
into a factor loading matrix $B$ of size $|\omega| \times f$ and a diagonal term $C=\text{diag}(c_i)$.
While the $c_i$ are the standard deviations of an independent multinormal surrogate, $B$ allows a covariance structure of rank $f$, while keeping the total number of variational parameters low at $|\theta|=|\omega| (f+2)$.
Although $f=1$ would allow a closed form of the natural gradients \citep{tran2020bayesian}, we used traditional backpropagation to experiment with $f\in\{0,1,2\}$, without computational limitations.

The underlying neural network uses batch normalization (BN; \citealt{ioffe2015batch}), which maintains a running average of means and variances that is  updated with each mini-batch during training.
Although \citet{mukhoti2020batch} derived that BN does not interfere with VI techniques, we found deteriorated performance and an increase in noise when using them in combination, and therefore we disabled all BN layers for the VI methods.

As a parameter prior, we set an isotropic normal distribution with dimension-independent variance $\sigma^2$ and trained the model both with fixed and variable $\sigma$.
As pointed out in \citet{murphy2023probabilistic}, this procedure breaks the Bayesian idea, but it can be useful in real applications.
Because the prior acts as a regularization, this construction allows the model to self-regularize. In fact, no overfitting is observed.
To stay data-agnostic, we did not set a predictive prior $p(y)$ for this work, but choices based on astronomical object spatial density, distance and signal strength, or catalogue labels are conceivable.  
For the update we used a stochastic scheme (SVI; \citealt{hoffman2013stochastic}), where in each gradient descent step only one mini-batch is consumed for the likelihood approximation.
To maintain balance within the ELBO, this requires the rescaling of the likelihood because it scales with the number of data points, while the KL terms scale with the number of model parameters. 
For the computation of the expectation values in Eq. \ref{eq:vi} we used three MC steps per gradient descent step. We find this choice to have little effect on the final training state, but more effect on the noise level of the ELBO value.

In each update we therefore sample three times from the parameter distribution and evaluate the ELBO on a random batch of size 128. We define an epoch as the number of updates equivalent to using the full set of training spectra: $updates\_per\_epoch=n_{train}/(n_{samples} \cdot batch\_size)$.

\section{Results} \label{sec:results}

\begin{table*}[t!]
\caption{
        Results of all models used for the SDSS dataset.  
    }
    \centering
    \begin{tabular}{ |p{3.4cm}||p{1.7cm}|p{1.9cm}|p{1.5cm}|p{1.9cm}|p{1.7cm}|p{1.3cm}|p{1.4cm}|  }
        \hline
        \multicolumn{8}{|c|}{Results (SDSS dataset)} \\
        \hline
        Model                                & Accuracy [$\%$]      & AUC $\times 1000$   & ECE [$\%$]  & NLL     & Brier score $\times 1000$    & Training time [s] & Inference time [ms]  \\
        \hline
Baseline CNN                         & $91.52 \pm 0.02$     & $997.4\pm0.2$       & $\bm{1.78 \pm 0.05}$& N/A                  & N/A              &  $\bm{95}$ & $\bm{0.012}$ \\
Dirichlet                            & $87.9 \pm 0.4$       & $988\pm1$           & $5.47 \pm 0.53$       & $0.76\pm0.07$        &$31\pm3$          &  $\bm{223}$ & $0.029$ \\
MCD100 ($p=0.125$)                   & $\bm{92.6 \pm 0.1}$  & $\bm{998.07\pm0.01}$& $\bm{1.80 \pm 0.02}$& $\bm{0.210\pm0.003}$ &$10.34\pm0.03$    &  $\bm{436}$ & $0.852$ \\
MCD1000 ($p=0.125$)                  & $\bm{92.58 \pm 0.07}$& $997.98\pm0.01$     & $\bm{1.80 \pm 0.02}$& $\bm{0.206\pm0.009}$ &$\bm{9.79\pm0.02}$&  $\bm{436}$ & $8.479$ \\
VI ($f=0$,$\sigma=0.3$)              & $91.35\pm 0.07$ & $997.35\pm 0.04$ & $2.21\pm0.01$ & $0.230\pm0.001$ & $11.34\pm 0.03$ & $1365$ & $0.609$ \\
VI ($f=1$,$\sigma=0.3$)              & $91.30\pm 0.10$ & $997.27\pm 0.06$ & $2.24\pm0.02$ & $0.233\pm0.003$ & $11.44\pm 0.11$ & $2175$ & $0.617$ \\
VI ($f=2$,$\sigma=0.3$)              & $91.15\pm 0.15$ & $997.22\pm 0.03$ & $2.25\pm0.02$ & $0.236\pm0.002$ & $11.50\pm 0.08$ & $2135$ & $0.616$ \\
VI ($f=0$,$\sigma=\text{variable}$)  & $89.22\pm 0.22$ & $995.83\pm 0.14$ & $2.86\pm0.05$ & $0.296\pm0.005$ & $13.97\pm 0.22$ & $1759$ & $0.620$ \\
VI ($f=1$,$\sigma=\text{variable}$)  & $89.47\pm 0.75$ & $996.00\pm 0.44$ & $2.76\pm0.12$ & $0.288\pm0.015$ & $13.54\pm 0.64$ & $2318$ & $0.620$ \\
VI ($f=2$,$\sigma=\text{variable}$)  & $89.48\pm 0.26$ & $996.07\pm 0.19$ & $2.71\pm0.05$ & $0.284\pm0.006$ & $13.31\pm 0.25$ & $2537$ & $0.617$ \\
        \hline
    \end{tabular}
    \tablefoot{
       Shown are the means and standard deviations of three independent training experiments with the same data splits, but different initialization weights.
        The highlighted values are the best in the column, within their uncertainty.
        $p$, $f$, and $\sigma$ denote MC dropout rate, the rank of the VI covariance, and the standard deviation of its parameter prior, respectively. 
        Training times are given until early stopping; inference times are given per spectrum. Both were taken on an Nvidia A5000 GPU.
    }
    \label{tab:results}
\end{table*}

\begin{figure}[h!]
    \centering
    \begin{subfigure}{\hsize}
        \includegraphics[width=\hsize]{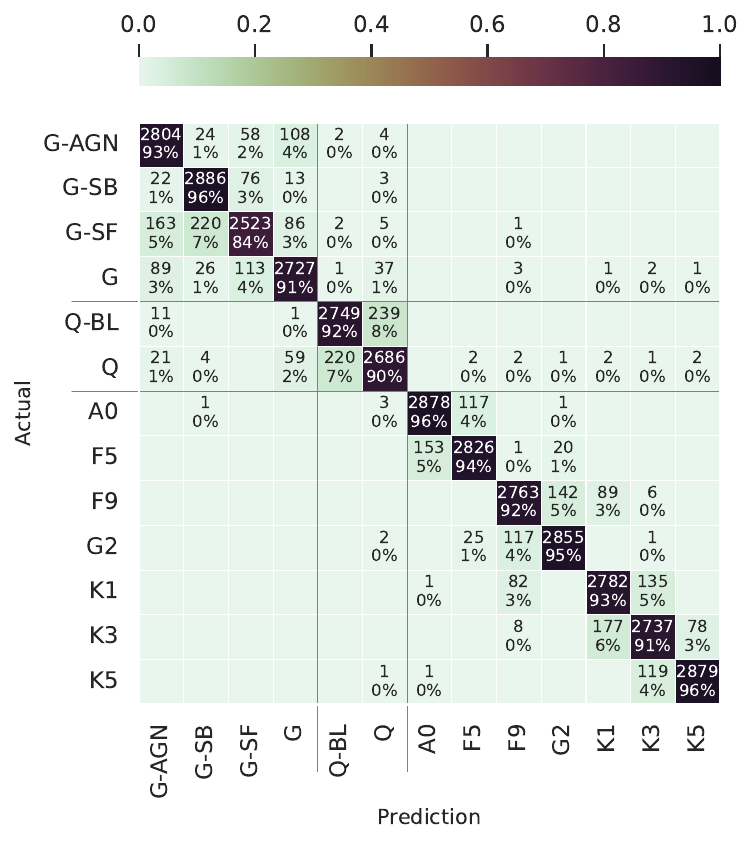}
    \end{subfigure}
    \caption{Confusion matrix for a Monte Carlo dropout model on the SDSS dataset. The class names were abbreviated. The colours indicate percentages. The black lines separate the classes of different 4MOST coarse labels: star, quasar, galaxy. The percentages may not sum to exactly 100\% due to rounding.}
    \label{fig:confusion}
\end{figure}

\begin{figure}[h]
    \centering
    \begin{subfigure}{\hsize}
        \includegraphics[width=\hsize]{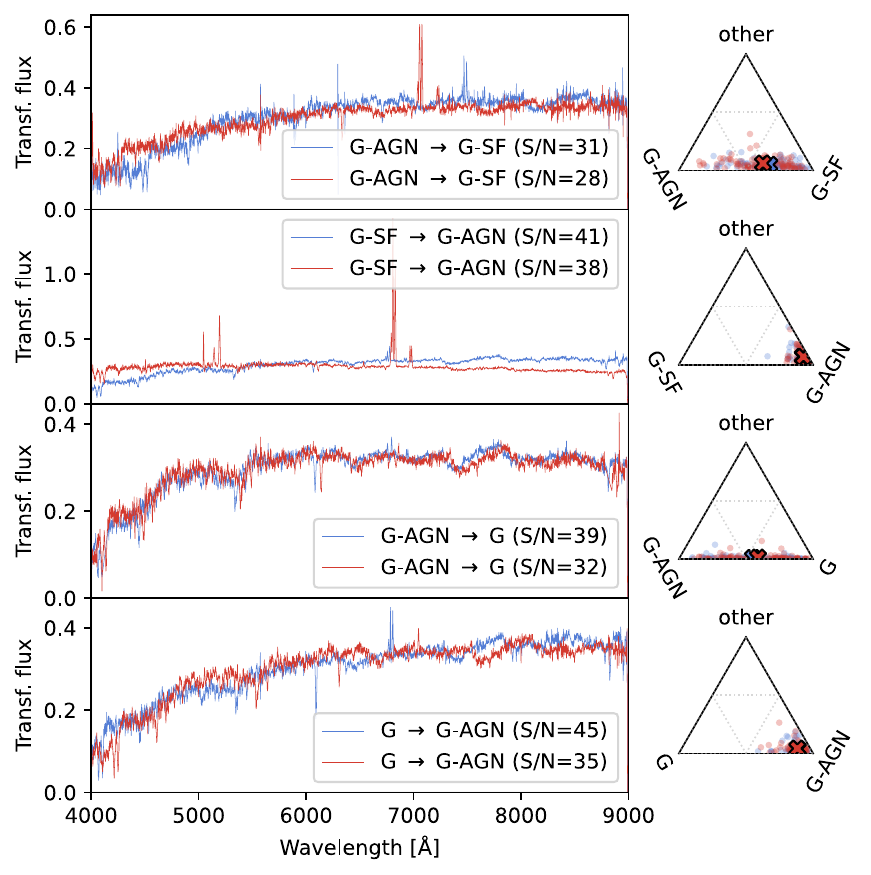}
    \end{subfigure}\\
    \caption{Examples for some of the most common confusion pairs on the SDSS dataset. Left: Two highest S/N spectra. The label format is ($\text{actual}\rightarrow\text{prediction}$). 
    Right: 100 samples from the MCD predictive posterior, where all other classes were marginalized into `other'. The crosses show the expectation values. }
    \label{fig:mcd_spectra_confusion}
\end{figure}

\begin{figure*}[t!]
    \centering
    \begin{subfigure}{0.33\hsize}
        \includegraphics[width=\hsize]{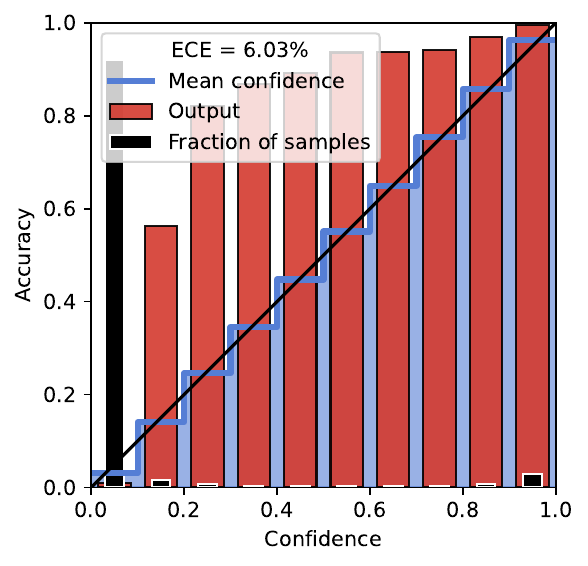}
    \end{subfigure}
    \begin{subfigure}{0.33\hsize}
        \includegraphics[width=\hsize]{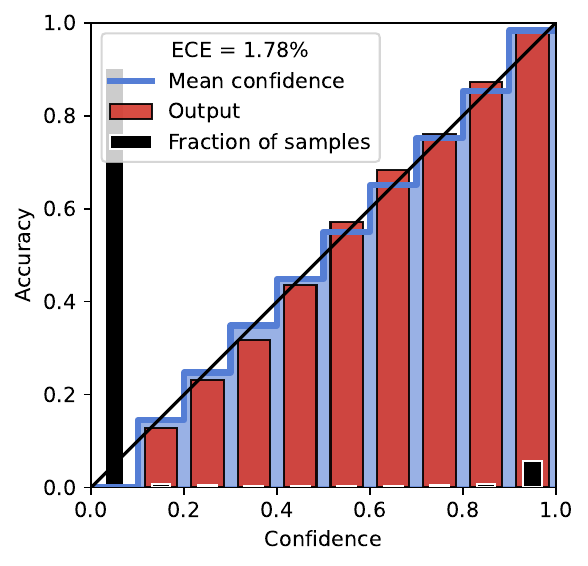}
    \end{subfigure}
    \begin{subfigure}{0.33\hsize}
        \includegraphics[width=\hsize]{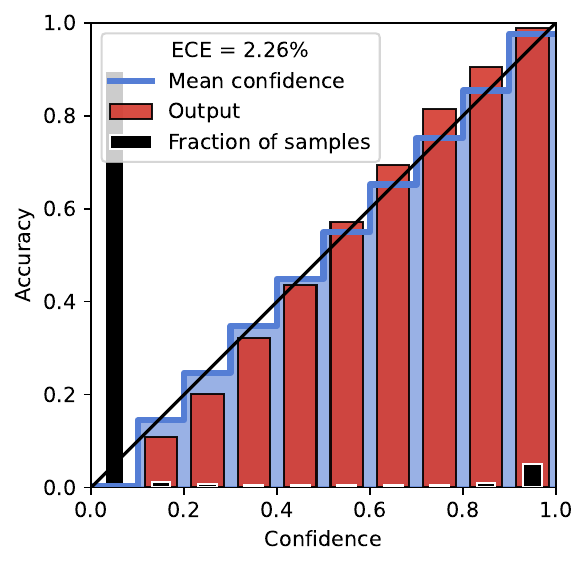}
    \end{subfigure}
    \caption{
        Calibration diagrams (also known as reliability diagrams) for three models: Dirichlet (left), MCD (centre), and VI with rank $f=1$ and fixed prior (right).
        For each input $(x,y)$ and each class $c$, the blue bars indicate how often $c$ coincides with $y$, given the binned predicted probability $p(y|x)$.
        The perfect values are shown in red, and take the binning effect into account. 
        The ECE is the expected calibration error and measures the deviation between the blue bars and the red line, weighted by the respective prediction rate (black). A lower ECE is better.
        Blue bars above and below the red line correspond to under- and overconfidence of the model, respectively.
        These plots average all the probabilities and the error over all $c$.
    }
    \label{fig:calibration}
\end{figure*}

In this section we compare the four methods in terms of their discriminative performance, as well as their ability to provide meaningful uncertainty estimates.
For each model we performed three independent and unbiased training experiments that started with different randomly initialized weights. We did not hand-pick the presented training runs. All runs of all models used the same split of data into training, validation, and test sets.
All the calculated metrics are tabulated in \Cref{tab:results,tab:mock_results}.
All the following figures are based on evaluating the models on the unseen test set.

The training histories of all four models on the 4MOST dataset are shown in Fig. \ref{fig:training}.
While the CNN exhibits clear overfitting beyond epoch 30, the probabilistic models show significantly reduced overfitting;  the gap between training and validation accuracy remains below 0.5 percentage points.
Particularly, the MC dropout makes the validation accuracy even slightly exceed the training accuracy.
All three runs converge in a similar manner for all models, which shows that the training is stable and converges reliably to the same posterior.
The Dirichlet model displays noticeably larger accuracy fluctuations across epochs, despite the use of low learning rates.
The training curves for the SDSS dataset are qualitatively identical, and are therefore omitted for brevity.

The reference CNN achieves a 13-way classification accuracy of $91.5\%$  on the SDSS dataset and a 10-way accuracy of $92.8\%$ on the mock dataset.
For comparison, the much larger ResNet-like network from \citet{zhong2023galaxy} achieves respective accuracies of $92.4\%$ and $93.4\%$ on the same datasets.
The Dirichlet model underperforms in all metrics, which is probably due to its artificially enforced form of the posterior. 
The VI models show worse accuracy than the baseline CNN;  allowing a dependence between the posteriors of individual weights ($k>0$) shows no improvement, thus indicating that these models are not limited by their fixed form parameter posterior $q_\theta$. Instead, the limiting factor might be the choice of the prior. 
The MCD model with best-fitting hyperparameters outperforms the baseline CNN in terms of accuracy, even though probabilistic models are generally slightly inferior in this regard \citep{gal2016uncertainty, gawlikowski2023survey}.
For all samplers we chose a validation MC sample size of 100. No performance gain is observed for higher sample numbers, but the inference time scales linearly with the number of samples.
For reference, an MCD inference run with a sampling size of 1000 is included in \Cref{tab:results}.

\begin{figure}[ht]
    \centering
    \includegraphics[width=\hsize]{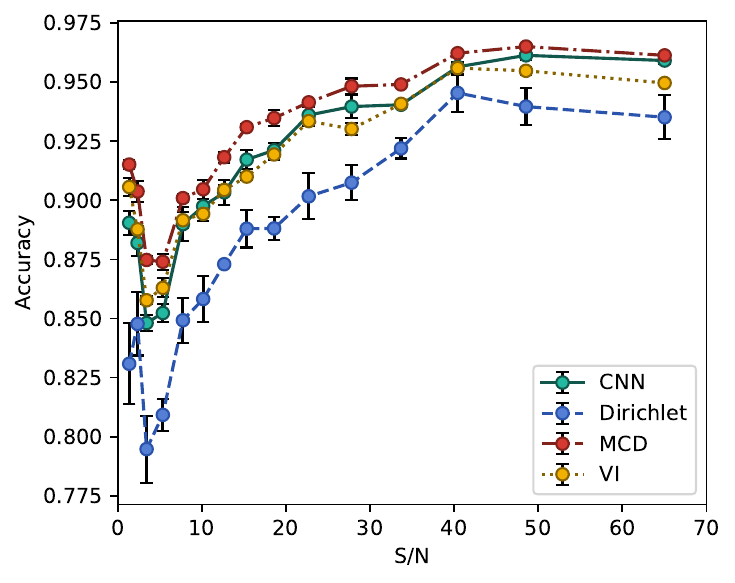}
    \caption{
        Effect of S/N on accuracy for all models. The error bars indicate standard deviations from three independent training experiments (SDSS dataset).
    }
    \label{fig:snr_accuracies}
\end{figure}

The confusion matrix for the SDSS test set (Fig. \ref{fig:confusion}) summarizes the counts of the true versus the predicted labels across all classes.
Overall, the model distinguishes well between galactic and extragalactic objects, with almost no confusion observed between these coarse classes.
Within the stellar classes, about 80\% of the misclassifications occur between adjacent spectral types.
This is expected, as the Morgan–Keenan spectral classification scheme underlying the labels is based on a continuous temperature sequence with boundaries that do not correspond to distinct spectral features.
This smooth gradation in stellar spectra naturally leads to overlap in feature space, particularly between neighbouring types.
In contrast, the separation between stars and quasi-stellar objects (QSOs) is remarkably clean despite the historical terminology.
Within the extragalactic classes, QSOs are well separated from other galaxies, with minimal cross-class confusion.
However, their broad-line nature is misclassified in approximately seven percent of the cases, indicating some ambiguity in the identification of spectral line widths.
Starburst (SB) galaxies are primarily confused with star-forming (SF) galaxies, which is expected given their similar emission-line features and overlapping star formation indicators.
In contrast, the SF class exhibits substantial confusion with all other extragalactic classes, reflecting its broad spectral diversity and overlap with both AGN-hosting and quiescent galaxies.
The confusion between AGN and normal galaxies lacks a clear physical interpretation. This may point to limitations in the spectral resolution or feature representation used by the model, or possibly to an intrinsic ambiguity in the labelling of weak or composite AGN spectra.
A closer inspection reveals that many actual--prediction pairs are in fact confused due to low S/N (Fig. \ref{fig:mcd_confusion_snr}).

Example spectra of commonly misclassified extragalactic sources are shown in Fig. \ref{fig:mcd_spectra_confusion}. The ternary distribution plots on the right give an impression of the models certainty. While there is high confidence in the wrong AGN predictions (rows two and four), labelled AGN are classified with significant uncertainty (rows one and three). We note that AGN and galaxies with star-forming activity are often confused due to their physical mechanisms and overlapping definitions \citep{teimoorinia2024revisiting}.

\begin{figure}[ht]
    \centering
    \begin{subfigure}{\hsize}
        \includegraphics[width=\hsize]{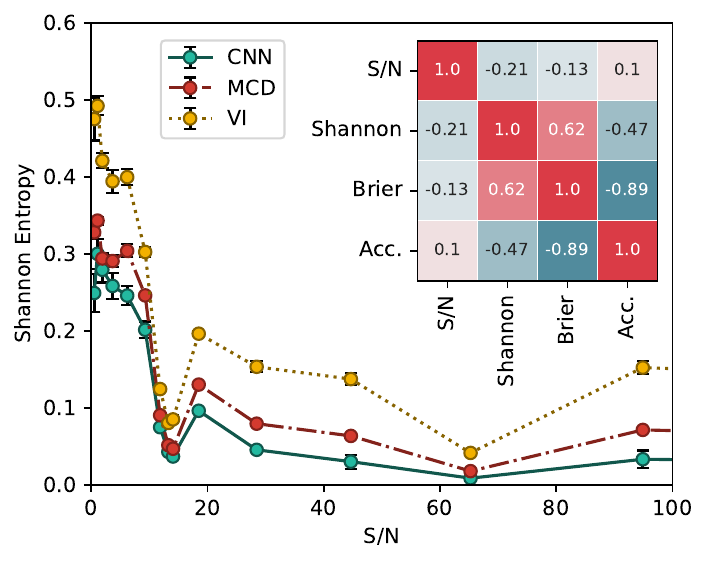}
    \end{subfigure}\\
    \caption{Shannon entropy vs S/N on the mock dataset. The Dirichlet model predictions have much higher entropy. The correlation matrix shows the relation of S/N (cause), accuracy (effect), Brier score (measure), and Shannon entropy (indicating uncertainty) of MCD predictions.}
    \label{fig:mock_snr_shannonEntropies}
\end{figure}

\begin{table*}[ht]
\caption{
        Results for the 4MOST mock dataset.     }
    \centering
    \begin{tabular}{ |p{3.4cm}||p{1.7cm}|p{1.9cm}|p{1.5cm}|p{1.9cm}|p{1.7cm}|p{1.2cm}|p{1.4cm}|  }
        \hline
        \multicolumn{8}{|c|}{Results (4MOST mock dataset)} \\
        \hline
        Model                                & Accuracy [$\%$]      & AUC $\times 1000$   & ECE [$\%$]  & NLL     & Brier score $\times 1000$    & Training time [s] & Inference time [ms]  \\
        \hline
Baseline CNN                          & $92.80\pm 0.08$      & $995.15\pm 0.04$ & $\bm{1.82\pm0.01}$      & N/A             &  N/A            & $ \bm{229}$ & $\bm{0.015}$ \\
Dirichlet                             & $87.37\pm 0.09$      & $986.95\pm 0.72$ & $6.01\pm0.16$      & $0.620\pm0.013$ & $38.71\pm 0.99$ & $ 968$ & $0.037$ \\
MCD50 ($p=0.125$)                     & $\bm{93.94\pm 0.07}$ & $\bm{996.25\pm 0.02}$ & $1.87\pm0.01$ & $\bm{0.142\pm0.001}$ & $ \bm{9.59\pm 0.05}$ & $ 576$ & $0.581$ \\
MCD100 ($p=0.125$)                    & $\bm{93.95\pm 0.08}$ & $\bm{996.25\pm 0.02}$ & $1.87\pm0.01$ & $\bm{0.142\pm0.001}$ & $ \bm{9.60\pm 0.05}$ & $ 576$ & $1.168$ \\
MCD200 ($p=0.125$)                    & $\bm{93.95\pm 0.05}$ & $\bm{996.25\pm 0.02}$ & $1.87\pm0.01$ & $\bm{0.142\pm0.001}$ & $ \bm{9.60\pm 0.05}$ & $ 576$ & $2.336$ \\
VI ($f=0$,$\sigma=0.3$)               & $91.71\pm 0.14$      & $994.71\pm 0.17$ & $2.60\pm0.05$      & $0.192\pm0.004$ & $13.26\pm 0.21$ & $1124$ & $0.799$ \\
VI ($f=1$,$\sigma=0.3$)               & $91.73\pm 0.38$      & $994.72\pm 0.12$ & $2.59\pm0.07$      & $0.191\pm0.005$ & $13.15\pm 0.36$ & $1422$ & $0.807$ \\
VI ($f=2$,$\sigma=0.3$)               & $91.43\pm 0.17$      & $994.50\pm 0.11$ & $2.69\pm0.01$      & $0.199\pm0.002$ & $13.67\pm 0.11$ & $1436$ & $0.804$ \\
VI ($f=0$,$\sigma=\text{variable}$)   & $89.14\pm 0.77$      & $992.37\pm 0.81$ & $3.48\pm0.30$      & $0.260\pm0.020$ & $16.95\pm 1.23$ & $ 665$ & $0.810$ \\
VI ($f=1$,$\sigma=\text{variable}$)   & $88.74\pm 0.55$      & $992.02\pm 0.82$ & $3.67\pm0.34$      & $0.272\pm0.024$ & $17.86\pm 1.51$ & $ 988$ & $0.809$ \\
VI ($f=2$,$\sigma=\text{variable}$)   & $89.55\pm 0.85$      & $992.45\pm 0.62$ & $3.46\pm0.26$      & $0.258\pm0.018$ & $16.81\pm 0.98$ & $ 959$ & $0.805$ \\
        \hline
    \end{tabular}
    \tablefoot{All values, as in \Cref{tab:results}.}
    \label{tab:mock_results}
\end{table*}

In our evaluation, no model outperformed the baseline CNN in terms of probability calibration. The expected calibration error (ECE) remains low ($\leq3\%$) across all models, with the exception of the Dirichlet model, which exhibits noticeably worse calibration.
Figure \ref{fig:calibration} displays three representative reliability diagrams. The subpar calibration performance of the Dirichlet model is likely attributable to its limited degrees of freedom, which constrain its ability to capture complex predictive uncertainty. This limitation could potentially be addressed by replacing the Dirichlet distribution with a more expressive alternative, such as a normalizing flow \citep{rezende2015variational}.

The performance of classifiers is expected to depend on the signal quality.
Figure \ref{fig:snr_accuracies} shows how the accuracy of all models correlates positively with S/N and levels out above $\text{S/N}\approx 40$. For these low-noise spectra, MCD reaches accuracies of over 96\%.
The lowest performance is obtained at $\text{S/N}\approx5$ and increases again for high-noise signals $\text{S/N}\leq5$, due to the reduced data variety in accordance with the Gini impurity (cf. Fig. \ref{fig:data_entropy}).
This behaviour is observed for all four models, while their ordering in accuracy is consistent across signal strengths. 
Comparison to the distribution of S/N in the dataset suggests that the total accuracy is limited by the noise in the extragalactic targets. Application within 4MOST may therefore yield higher (or lower) values, while MCD is expected to consistently perform best on other data.

To provide an estimate of how well the algorithms work on an unbiased sample of the sky (rather than on a class-balanced test set), we first estimated the ratio of galactic to extragalactic sources from the photometric SDSS table ($PhotoObjAll$) to be $1.37$, and we then used the subclass counts from the spectroscopic table ($SpecObj$) within each coarse class. The resulting class ratios are dominated by galaxies (27\%) and F stars (31\%), while AGN and starburst galaxies are least abundant (together 1.2\%). Using these values as weights to the per-class accuracy of the trained MCD models results in an adjusted total accuracy of $92.0\% \pm 0.3 \%$. This value is slightly lower than on the balanced test set;  galactic sources have a positive impact on the difference, while extragalactic contribute negatively.

\begin{figure}[h!]
    \centering
    \begin{subfigure}{\hsize}
        \includegraphics[width=\hsize]{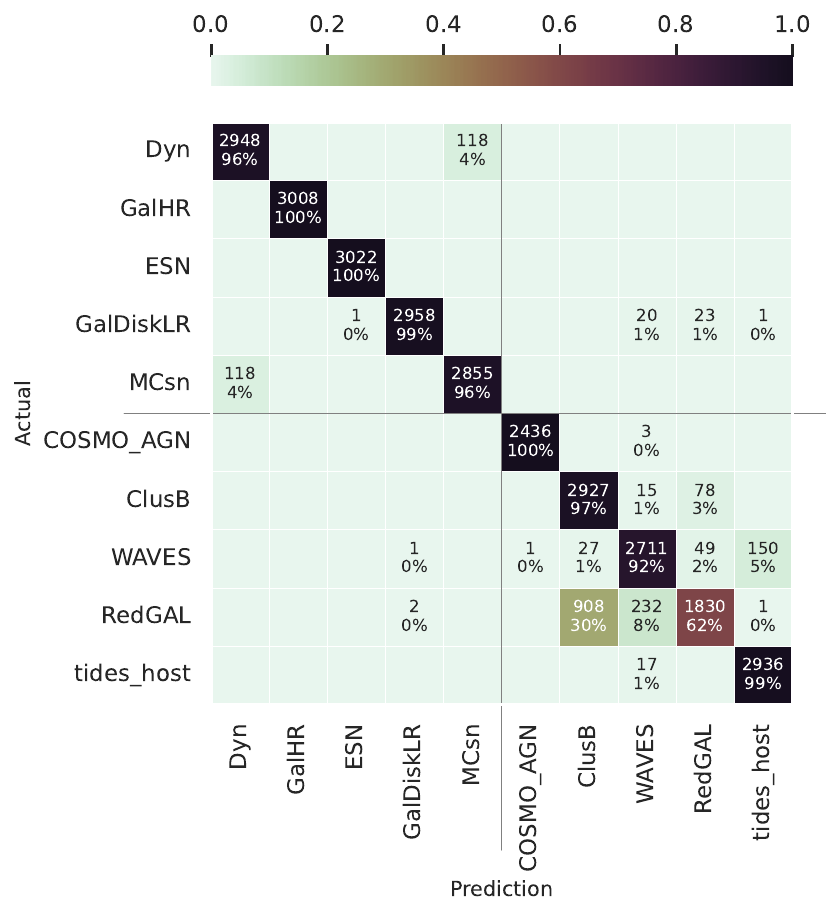}
    \end{subfigure}%
    \caption{Class confusion of the MCD model for the 4MOST mock dataset, as in Fig. \ref{fig:confusion}. The black lines separate blocks of extragalactic and galactic classes.}
    \label{fig:mock_confusion}
\end{figure}

The NLL and Brier score are not themselves measures of uncertainty, but are indicators of the calibration of uncertainty, given that the true label is known \citep{lakshminarayanan2017simple, gal2016uncertainty}.
Both metrics reward high-confidence correct predictions and heavily penalize confident errors.
Their strong threshold-like separation (Fig. \ref{fig:uncertainties}) between correct and incorrect predictions for the Bayesian methods reflects how the two metrics respond to the confidence encoded in the predictive distribution, and indicates that the models' predicted probabilities are well aligned with the actual outcomes, such that confidence is a reliable proxy for correctness.
The strong similarity of the distributions of MCD and VI suggests that both models are learning the same posterior, while MCD is better able to resolve an off-Gaussian PDF.
The fact that the distribution looks so different for the Dirichlet model may be another hint of the lack of the model's flexibility, as mentioned above.

A statistical statement on the epistemic uncertainties, delivered by the probabilistic models is made in Fig. \ref{fig:mock_snr_shannonEntropies}.
While the drops in prediction entropy at $S/N\approx14$ and $S/N\approx65$ can again be attributed to the low data entropy (Fig. \ref{fig:mock_data_entropy}), we verified that the uncertainty correlates positively with the Brier score, but negatively with accuracy and signal quality, as confidence in incorrect predictions and predictions on high-noise signals is expected to decrease.

All the discussed differences between the models directly translate to the 4MOST mock dataset. 
Again, the MCD variants prevail as the models with highest accuracy and best uncertainties, independently of the number of used evaluation samples. 
The confusion matrix of the MCD model on the mock dataset (Fig. \ref{fig:mock_confusion}) shows very high accuracy for all galactic classes and AGN, probably due to the specificness of these classes within the class set. 
The 30\% misclassification of red galaxies as bright clusters can be explained by the cross-contamination of these classes in the training data \citep{zhong2023galaxy}.
Finally, a set of one-versus-rest ROC curves is shown in Fig. \ref{fig:mcd_roc}.

\FloatBarrier
\section{Conclusions}

We have shown that Bayesian approximate inference with Monte Carlo dropout is able to keep up with and even exceed the reference neural network in its predictive performance of the $K$-classification of astronomical spectra.
In addition, the model provides well-calibrated predictive posteriors on the probability simplex, while the associated increase in inference time for 100 samples, relative to a standard CNN, appears to be a negligible investment. 

Such a model could be suitable for the 4MOST classification pipeline, potentially applied to a more powerful deep learning architecture.
In this set-up condensed class-wise uncertainties could be easily obtained by marginalizing the predictive posterior for each class and compute  its standard deviation (assuming normal marginal), its two-sided percentiles, or its differential entropy.
These class uncertainties could then be used to guide object selection and quality assessment in downstream tasks. Especially when a large number of objects makes manual verification unfeasible, Bayesian uncertainties allow the quantification of errors that come with using misclassified objects. 
This way, a classifier with uncertainties could even improve the effective completeness of a 4MOST survey catalogue by correcting its labels and enabling the controlled inclusion of objects that would otherwise be excluded by conservative selection cuts, while keeping the class contamination quantifiable.
The related chosen uncertainty thresholds are survey and task specific, and are dependent on the object abundance, the desired number of usable objects, and the confused classes themselves.

On the contrary, using a neural network to predict a Dirichlet distribution to describe the posterior is clearly affected by the rigid form of the distribution. 
Although easy to adapt, this approach yields a poor accuracy and a poorly calibrated uncertainty.  
Despite their more principled Bayesian approach, the models trained with variational inference were found to be inferior in all metrics. Allowing a low-rank covariance had little effect on these results. This may be due to a bad choice of prior. Learning a diagonal normal prior during the training could not overcome these problems. 

\begin{acknowledgements}
    We thank  Thorsten Glüsenkamp and Anish Amarsi for  useful comments and fruitful discussions, and Fucheng Zhong for providing training data.
    We would like to acknowledge financial support from the eSSENCE graduate school for data intensive science. A.K. and M.S. were supported by the Swedish National Space Agency (SNSA).\\    
\end{acknowledgements}

\bibliographystyle{aa}
\bibliography{references.bib}

\begin{appendix}   

{
\section{4MOST mock dataset}

\begin{figure}[h]
    \centering
    \includegraphics[width=\hsize]{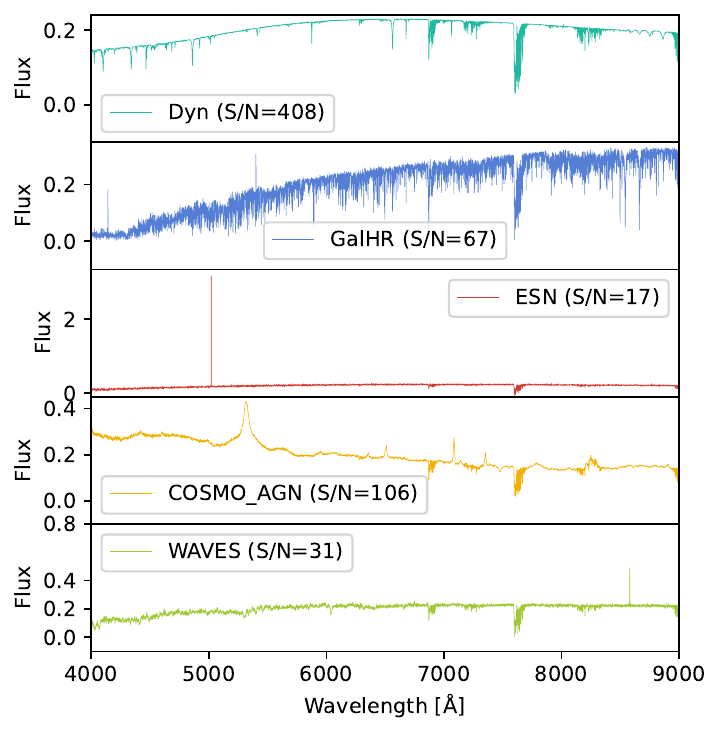}
    \caption{
        Selected example spectra of the mock dataset with the highest S/N of the respective class.
    }
    \label{fig:mock_data_spectra}
\end{figure}
\begin{figure}[h]
    \centering
    \includegraphics[width=\hsize]{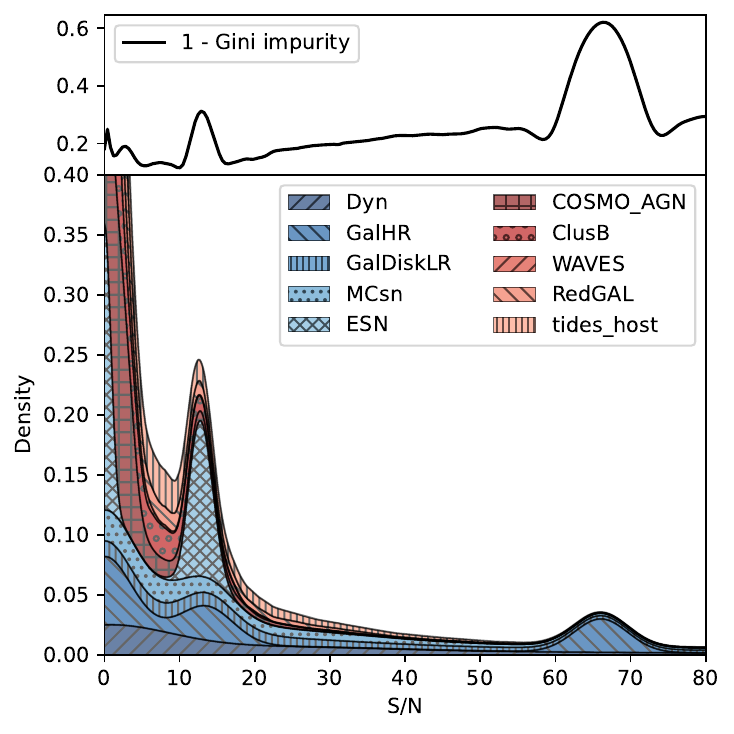}
    \caption{
        S/N distribution (bottom) and resulting expected random-guessing accuracy (top) as in Fig. \ref{fig:data_entropy} for the mock dataset.
    }
    \label{fig:mock_data_entropy}
\end{figure}

\section{Receiver operating characteristic}

In the multi-class setting, each ROC curve (Fig. \ref{fig:mcd_roc}) is computed by treating one class as the positive case and all the others as negative. This one-versus-rest approach isolates the model’s ability to distinguish each class from the remainder, independent of overall classification performance.
All models exhibit high area under the curve (AUC) values ($>0.99$), indicating excellent discriminative ability. Although overall accuracy is not perfect, the consistently high AUCs suggest that the models rank instances correctly with high confidence, even when the final, threshold-based prediction is incorrect. In many such cases, the true class still receives a relatively high predicted probability.
This implies that the models capture meaningful structure in the data and effectively encode class relationships in their output distributions.
In particular, a good ranking accounts for systematic label-related misclassifications, such as those between adjacent stellar spectral types.

\begin{figure}[t]
    \centering
    \begin{subfigure}{\hsize}
        \includegraphics[width=\hsize]{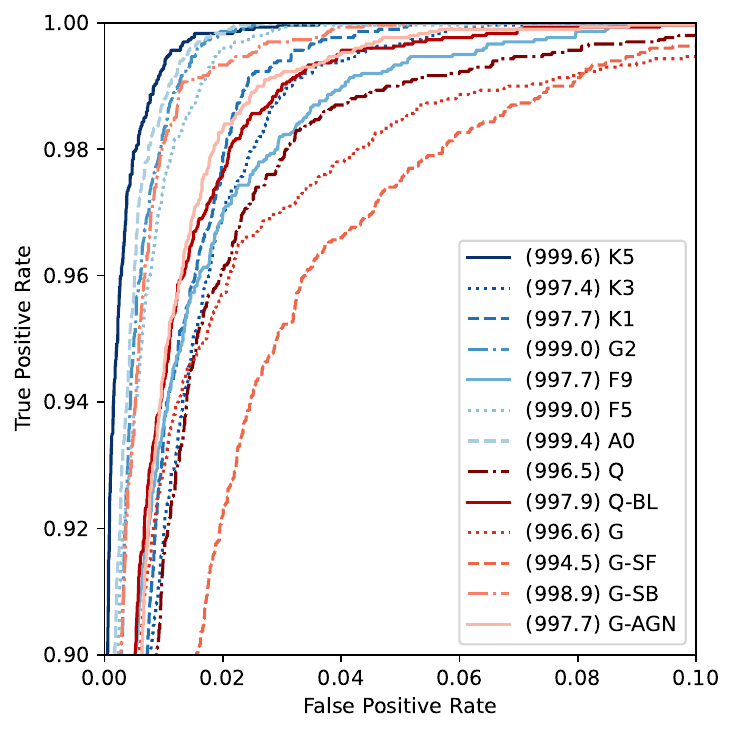}
    \end{subfigure}\\
    \caption{Receiver operator curve (ROC) of the MCD model on the SDSS dataset. In brackets $\text{AUC} \times 1000$. }
    \label{fig:mcd_roc}
\end{figure}

{
\section{Learning curves}

\begin{figure*}[t!]
    \centering
    \begin{subfigure}{\hsize}
        \includegraphics[width=\hsize]{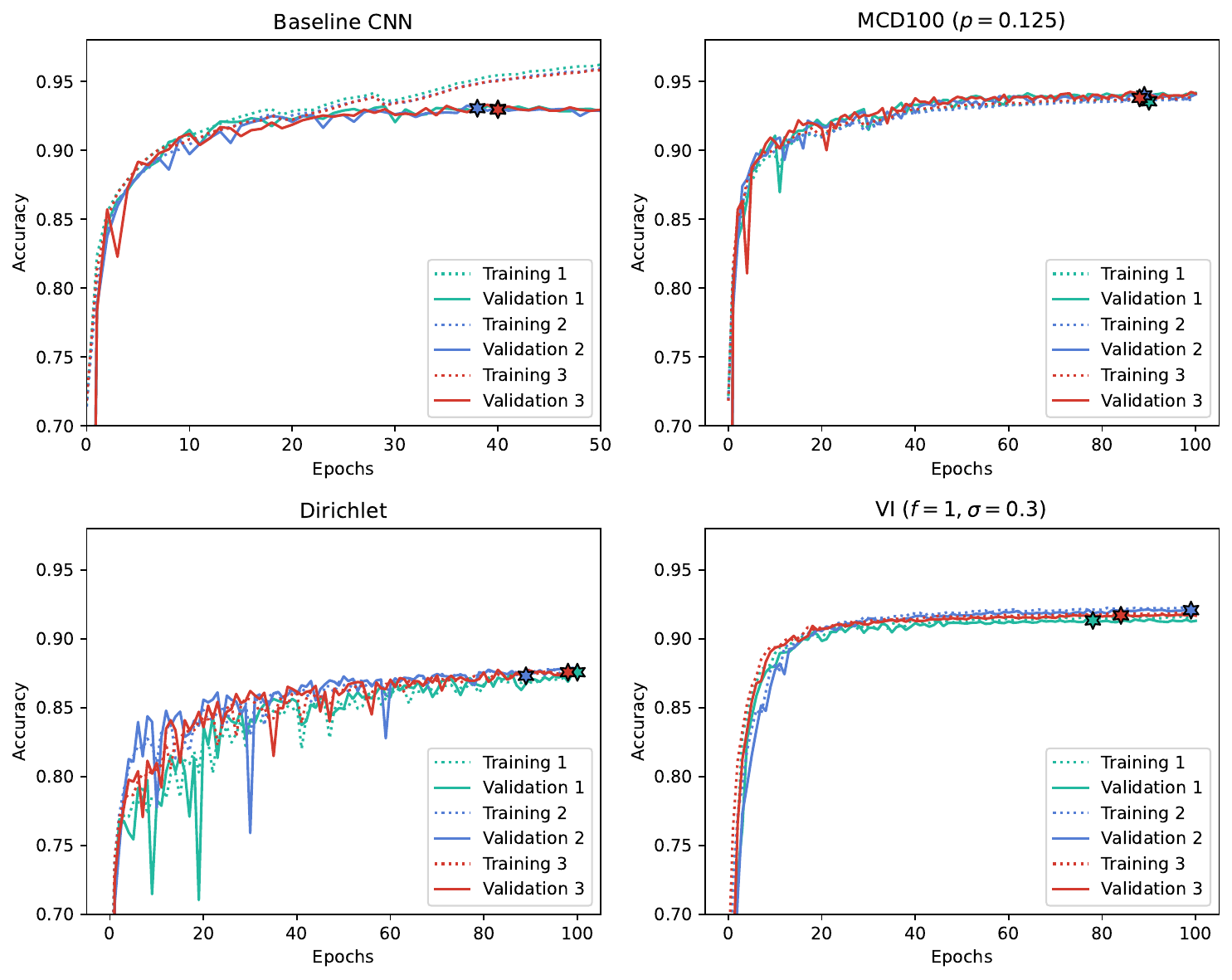}
    \end{subfigure}
    \caption{    
       Training curves for all models, showing the accuracies as a function of the number of trained epochs, for the three training runs on the 4MOST mock dataset. Loss functions are omitted for clarity. A strong separation between training and validation curves would indicate overfitting. A difference between runs would indicate unstable convergence and local loss minima.
        The stars indicate the training state with the lowest validation loss that  were used for the results above.  
    }
    \label{fig:training}
\end{figure*}
}

\section{S/N confusion}

Figure \ref{fig:mcd_confusion_snr} summarizes the highest S/N for each misclassification pair in both datasets.
These plots reveals to which extend certain classes are confused due to bad signal in contrast to other reasons. 
Especially the few off-block cases are thus completely explained by high noise. The dominant inter-block confusions, such as RedGAL $\rightarrow$ ClusB, on the other hand, may not be seen as caused by noise, given the generally low mean S/N of 8.7 for red galaxies.

\begin{figure*}[h]
    \centering
    \begin{subfigure}{0.498\hsize}
        \includegraphics[width=\hsize]{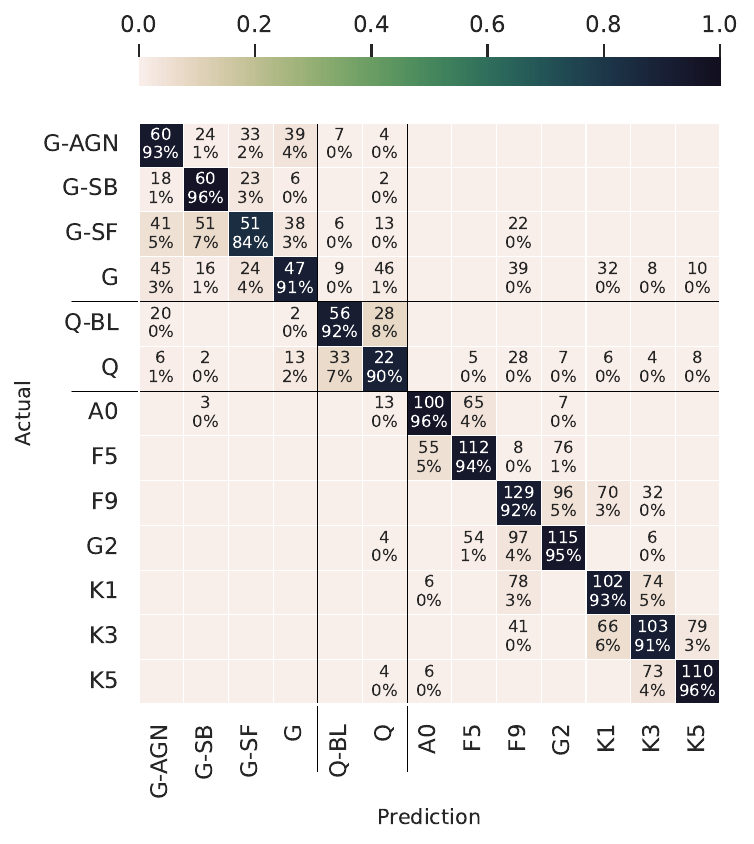}
    \end{subfigure}%
    \begin{subfigure}{0.502\hsize}
        \includegraphics[width=\hsize]{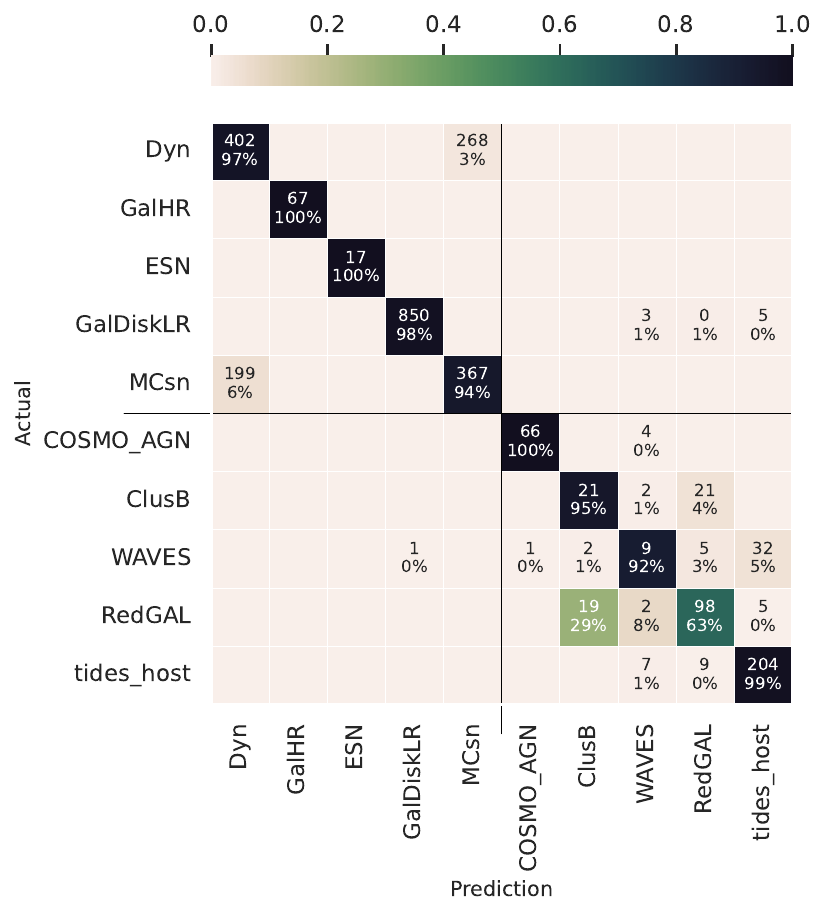}
    \end{subfigure}\\
    \caption{Confusion matrix. Top labels are the maximum S/N value for every actual--prediction pair. The bottom labels and colouring are as in  Fig. \ref{fig:confusion}. Left: SDSS dataset. Right: 4MOST mock dataset.}
    \label{fig:mcd_confusion_snr}
\end{figure*}

\section{Uncertainty distributions}

\begin{figure*}[t!]
    \centering
    \begin{subfigure}{0.95\hsize}
        \includegraphics[width=\hsize]{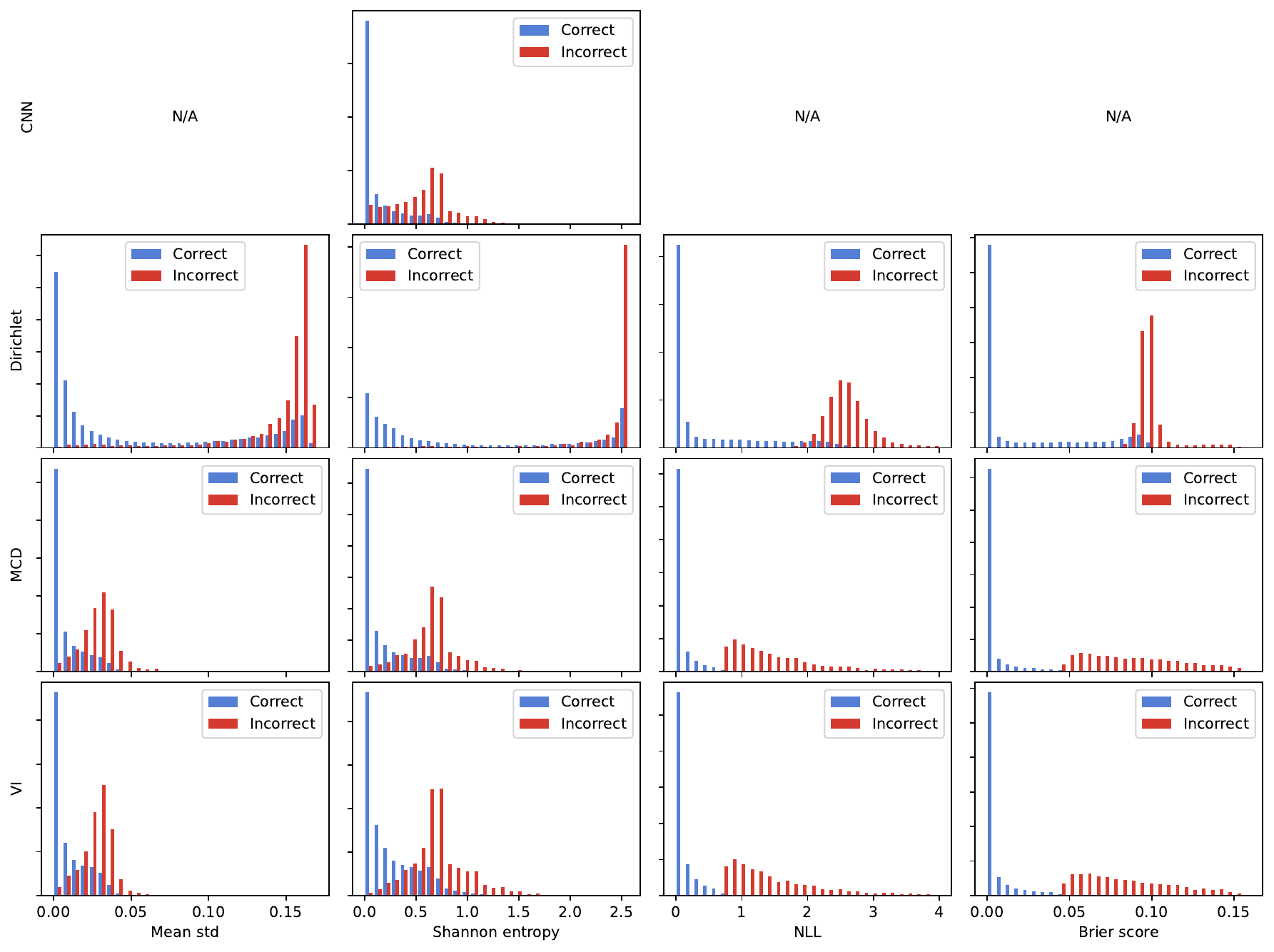}
    \end{subfigure}
    \caption{Comparison of different metrics of uncertainty on the SDSS dataset. Abscissa and ordinate labels apply to full columns and rows, respectively.  Correct and incorrect histogram bars are independently normalized to one.}
    \label{fig:uncertainties}
\end{figure*}

    Histograms of discussed uncertainty metrics are compared in Fig. \ref{fig:uncertainties}.
    All metrics correlate with prediction correctness (green vs red). 
}

\end{appendix}
      
\end{document}